\documentclass[aps,prl,english,superscriptaddress,twocolumn,tightenlines,showpacs,longbibliography,floatfix,amsfonts]{revtex4-1}
\usepackage[T1]{fontenc}
\usepackage[latin9]{inputenc}
\usepackage[normalem]{ulem}
\usepackage{textcomp}
\usepackage{graphicx,float}
\usepackage[dvipsnames,svgnames]{xcolor}
\usepackage{mathrsfs,mathtools,xfrac}
\usepackage{amsmath,amssymb,amsthm,amscd,bm,bbm}
\usepackage{txfonts} % better font, but put after ams math to avoid conflicts
\usepackage{afterpackage,multirow,comment}
\usepackage{array,booktabs,minibox}
	\newcolumntype{P}[1]{>{\centering\arraybackslash}p{#1}} % control the alignment and width
\usepackage[colorlinks,bookmarks=false,citecolor=blue,linkcolor=blue,urlcolor=blue,filecolor=blue]{hyperref}
\usepackage{wasysym}

\theoremstyle{remark}
\newtheorem*{claim*}{\protect\claimname}

\usepackage{babel}
\providecommand{\claimname}{Claim}

\newcommand{\mean}[1]{\left\langle #1 \right\rangle}

\newcolumntype{C}{>{$}c<{$}}
\AtBeginDocument{
\heavyrulewidth=.08em
\lightrulewidth=.05em
\cmidrulewidth=.03em
\belowrulesep=.65ex
\belowbottomsep=0pt
\aboverulesep=.4ex
\abovetopsep=0pt
\cmidrulesep=\doublerulesep
\cmidrulekern=.5em
\defaultaddspace=.5em
}

\begin{document}

\title{Exotic Symmetry Breaking Properties of Self-Dual Fracton Spin Models}

\author{Giovanni Canossa}
\affiliation{Department of Physics and Arnold Sommerfeld Center for Theoretical Physics, Ludwig-Maximilians-Universit\"{a}t M\"{u}nchen, Theresienstr.~37, D-80333 M\"{u}nchen, Germany}
\affiliation{Munich Center for Quantum Science and Technology (MCQST), Schellingstr.~4, D-80799 M\"{u}nchen, Germany}
\author{Lode Pollet}
\affiliation{Department of Physics and Arnold Sommerfeld Center for Theoretical Physics, Ludwig-Maximilians-Universit\"{a}t M\"{u}nchen, Theresienstr.~37, D-80333 M\"{u}nchen, Germany}
\affiliation{Munich Center for Quantum Science and Technology (MCQST), Schellingstr.~4, D-80799 M\"{u}nchen, Germany}
\author{Miguel A. Martin-Delgado}
\affiliation{Departamento de F\'isica Te\'orica, Universidad Complutense, 28040 Madrid, Spain}
\affiliation{CCS-Center for Computational Simulation, Universidad Polit\'ecnica de Madrid, 28660 Boadilla del Monte, Madrid, Spain}
\author{Hao Song}
\email{songhao@itp.ac.cn}
\affiliation{CAS Key Laboratory of Theoretical Physics, Institute of Theoretical Physics, Chinese Academy of Sciences, Beijing 100190, China}
\author{Ke Liu}
\email{ke.liu@ustc.edu.cn}
\affiliation{Hefei National Research Center for Physical Sciences at the Microscale and School of Physical Sciences, University of Science and Technology of China, Hefei 230026, China}
\affiliation{Shanghai Research Center for Quantum Science and CAS Center for Excellence in Quantum Information and Quantum Physics, University of Science and Technology of China, Shanghai 201315, China}
\affiliation{Department of Physics and Arnold Sommerfeld Center for Theoretical Physics, Ludwig-Maximilians-Universit\"{a}t M\"{u}nchen, Theresienstr.~37, D-80333 M\"{u}nchen, Germany}
\affiliation{Munich Center for Quantum Science and Technology (MCQST), Schellingstr.~4, D-80799 M\"{u}nchen, Germany}
\date{\today}

\begin{abstract}
Fracton codes host unconventional topological states of matter and are promising for fault-tolerant quantum computation due to their large coding space and strong resilience against decoherence and noise.
In this work, we investigate the ground-state properties and phase transitions of two prototypical self-dual fracton spin models---the tetrahedral Ising model and the fractal Ising model---which correspond to error-correction procedures for the representative fracton codes of type-I and type-II, the checkerboard code and the Haah's code, respectively, in the error-free limit.
They are endowed with exotic symmetry-breaking properties that contrast sharply with the spontaneous breaking of global symmetries and deconfinement transition of gauge theories.
To show these unconventional behaviors, which are associated with sub-dimensional symmetries, we construct and analyze the order parameters, correlators, and symmetry generators for both models.
Notably, the tetrahedral Ising model acquires an extended semi-local ordering moment, while the fractal Ising model fits into a polynomial ring representation and leads to a fractal order parameter.
Numerical studies combined with analytical tools show that both models experience a strong first-order phase transition with an anomalous $L^{-(D-1)}$ scaling, despite the fractal symmetry of the latter.
Our work provides new understanding of sub-dimensional symmetry breaking and makes an important step for studying quantum-error-correction properties of the checkerboard and Haah's codes.
\end{abstract}

\maketitle

\section{Introduction} \label{sec:intro}
A major task in the theoretical study of topological quantum computation is the search for the most resilient quantum error correction (QEC) codes.
Although the two-dimensional (2D) toric code~\cite{Kitaev03, Dennis02, Wang03} and color code~\cite{Bombin06, Katzgraber09, Bombin12} have a code capacity $p_c \simeq 10.9\%$ against random qubit errors and are ideal for topological quantum memory, they permit only transversal implementations of logical Clifford gates as any 2D topological stabilizer code~\cite{Bravyi13, Campbell17}.
Such gates can be efficiently simulated in polynomial time on classical computers~\cite{Aaronson04}. 
Therefore, despite the striking experimental advances of these codes on trapped-ion~\cite{Nigg14, Postler22, Ryan22} and superconducting-qubit~\cite{Satzinger21, Krinner22, Zhao22} platforms, in the long term, we have to go beyond 2D to realize the desired quantum advantages with topological protection.
However, in three dimensions, thresholds of standard topological stabilizer codes are significantly reduced by the multi-spin interactions and gauge symmetries in their error modeling.
For instance, the effective spin models determining the stability of 3D toric codes and color codes induce random $Z_2$ and $Z_2 \times Z_2$ gauge theories and exhibit optimal minimum thresholds $\simeq 3.3\%$~\cite{Wang03, Ohno2004} and $\simeq 1.9\%$~\cite{Kubica18} against random qubit errors, respectively. 
This hence motivates the quest for novel codes in three and higher dimensions.

Fracton codes represent a novel class of 3D topological stabilizer codes~\cite{Vijay16, Vijay15, Haah11, Bravyi13_RG}.
Such codes support subextensive ground-state degeneracies (GSDs) and can provide larger coding spaces than standard topological codes with constant degeneracy.
Moreover, their gapped excitations conform to intrinsic mobility constraints that can suppress error propagations and maintain the correctability of the codes.
Strong error resilience of fracton codes because of this latter feature is indeed confirmed in our recent work~\cite{Song22}.
There, we determined the error thresholds of the representative X-cube code with high accuracy and found that its code capacity, $p_c \simeq 7.5\%$, is remarkably higher than other 3D codes~\cite{Wang03, Ohno2004}, including 3D toric and color codes~\cite{Kubica18}.
 
In a broader sense, fracton systems constitute novel states of matter due to their unique symmetries and excitations.
Their unconventional many-body properties~\cite{Chamon05, Yoshida13, Ma17, Shirley18, Song19, Prem19, Slagle21, Prem17, Devakul19, Nandkishore19, Aasen20, Devakul18, Muhlhauser20, Zhou22, Zhu23, Canossa23, Rayhaun23, Zarei22, Yan20, Song24, Niggemann23} and field-theory formalisms~\cite{Seiberg21, Seiberg20, Seiberg21b, Pretko20, Bulmash18, Ma18, Ye20, Ye21, Beekman17, Pretko18, Yan19, Gromov24, Casasola23, Sfairopoulos23} are under active study in both condensed matter and high-energy physics, although requirements of an intrinsic topological order might be relaxed.

This work aims to carry out a necessary move toward understanding the QEC properties of the type-I checkerboard code and type-II Haah's code.
In Ref.~\cite{Song22}, we conjectured that these two self-dual fracton codes may reach the optimal error threshold $p_c \sim 11\%$ as a quantum memory.
Accurate determination of their thresholds requires constructing a disorder-temperature phase diagram for each model and demands exceeding computational resources owing to the interplay between fracton and spin-glass physics.
As a first step, we focus in the current work on the error-free limit of the two codes and investigate phase transitions and construct order parameters of their dual spin models.
Understanding the associated sub-dimensional symmetry breaking (SDSB) and {\it non-local} order parameters can guide future studies at finite error rates.

At the error-free limit, duals of the checkerboard and Haah's codes lead to the disorder-free tetrahedral Ising model (TIM) and fractal Ising model (FIM), respectively. 
Subextensive topological GSDs of the fracton codes are reflected by distinct sub-dimensional symmetries of the dual-spin models.
Such symmetries act on lower-dimensional manifolds or fractals of the system~\cite{note_subdim} and are intermediate between global and gauge symmetries.
Spontaneous symmetry breaking remains possible but produces long-range orders {\it without} a local order parameter.
The associated phase transitions fit neither the Landau paradigm nor the standard confinement-deconfinement scenario.
Their finite-size scalings also exhibit anomalous behaviors, despite both spin models investigated here experiencing a strong first-order transition.

The manuscript is organized as follows. In Sec.~\ref{sec:models}, we define the two fracton spin models, discuss their symmetry properties, and construct appropriate correlators and order parameters.
Sec.~\ref{sec:fractal_symmetry} is devoted to the fractal symmetry generators and the consequent degeneracy of the fractal Ising model, utilizing a polynomial ring formalism.
Sec.~\ref{sec:transition} discusses the phase transitions and scaling behaviors of the two models.
We conclude at Sec.~\ref{sec:sum} with an outlook.
App.~\ref{app:ring} and App.~\ref{app:simulation} include details of fractal generators and multi-canonical simulations.

\begin{figure}
  \centering
  \includegraphics[width=0.48\textwidth]{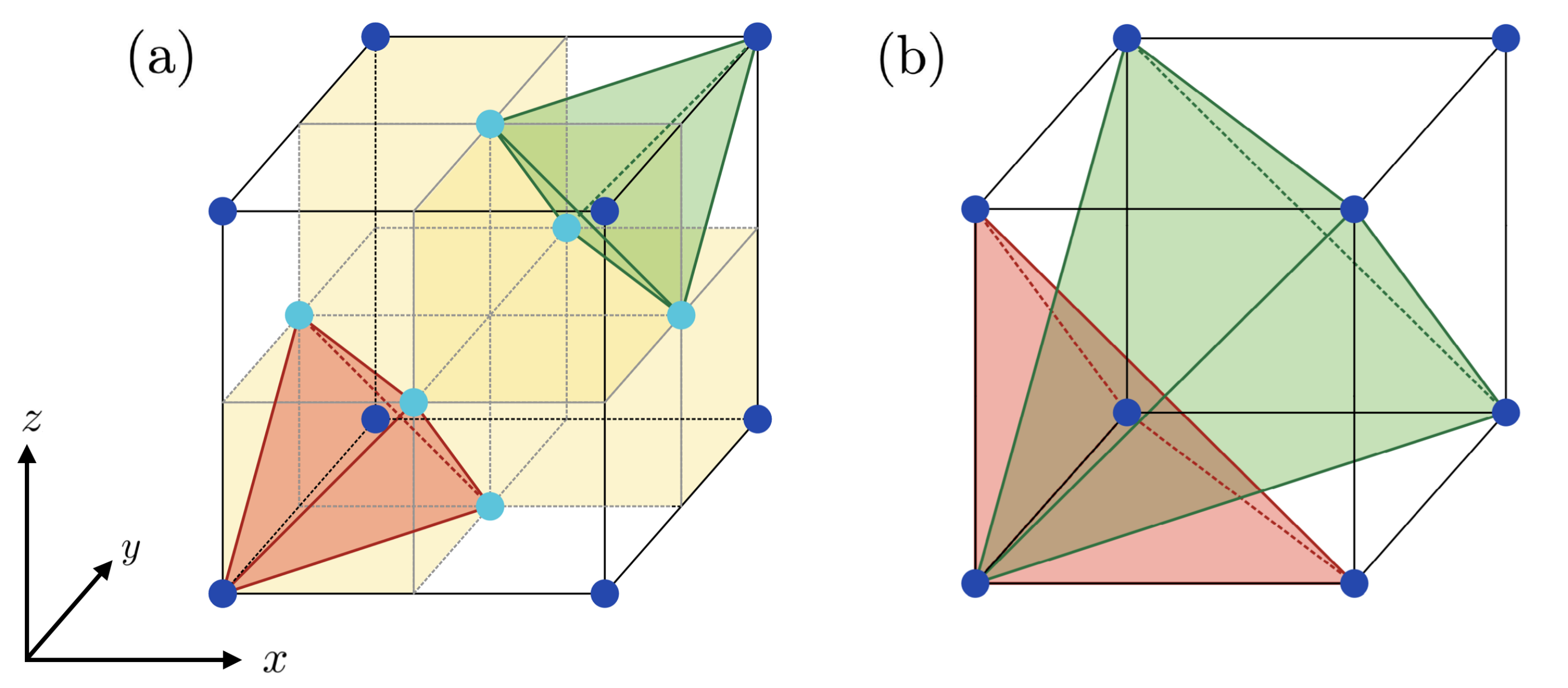}
  \caption{Illustration of the two fracton spin models. (a) A unit cell of the tetrahedral Ising model. Blue and cyan circles represent vertices ${\bf v} = (x, y, z) \in (2\mathbb{Z})^3$ and face centers of an FCC lattice, respectively. The red and green tetrahedra show an example of the $\prod_{\bf a} s_{{\bf v}+{\bf a}}$ and $\prod_{\bf a} s_{{\bf v}-{\bf a}}$ interaction terms, respectively, where ${\bf a} \in \left\{ (0,0,0), \, (1,1,0), \, (1,0,1), \, (0,1,1) \right\}$ labels the four FCC sublattices. The entire lattice can be intuitively visualized with small cubes: Each shaded (empty) cube contains a single red (green) tetrahedron interaction.
  		(b) A unit cell of the fractal Ising model on a simple cubic lattice. Vertices are again denoted by blue circles, but here ${\bf v} = (x, y, z) \in \mathbb{Z}^3$. The red and green tetrahedra represent the nearest-neighboring and next-nearest-neighboring interaction terms $\prod_{{\bf a}_1} s_{{\bf v}+{\bf a}_1}$ and $\prod_{{\bf a}_2} s_{{\bf v}+{\bf a}_2}$, respectively, with ${\bf a}_1 \in \left\{(0,0,0), \, (1,0,0), \, (0,1,0), \, (0,0,1)\right\}$ and ${\bf a}_2 \in \left\{(0,0,0), \, (1,1,0), \, (1,0,1), \, (0,1,1)\right\}$.}
  \label{fig:model}
\end{figure}

\section{Fracton spins models} \label{sec:models}
The tetrahedral and fractal Ising models relate to the checkerboard and Haah's codes in two ways.
One way is using a generalized Wegner's duality~\cite{Wegner71, Wegner73} where the two fracton codes are obtained by gauging these two spin models~\cite{Vijay16}.
In this correspondence, immobile excitations in a fracton code map to domain-wall corners of its dual spin model.
Furthermore, a fracton topological order is dual to an SDSB phase on the spin-model side.

The other way is through a statistical-mechanical mapping that describes the error-correction procedure of a topological code. 
This approach is initially developed for standard gauge codes, such as toric and color codes~\cite{Dennis02, Bombin12, Kubica18}, and has been recently extended to fracton codes~\cite{Song22}.
Under this mapping, a fracton spin model corresponds to the error-free limit of a fracton code, while qubit errors are reflected by including quenched disorders.

Here, we focus on the disorder-free limit, but will study the more involved disorder-full case in a future work.

\subsection{Tetrahedral Ising model} \label{sec:tetra-Ising}

The tetrahedral Ising model is dual to the checkerboard model. Its Hamiltonian $H_{\rm TIM}$ is defined on a face-centered cubic (FCC) lattice, with one Ising spin $s = \pm 1$ placed at each lattice site, as depicted in Fig.~\ref{fig:model}.
For convenience, we use ${\bf v} = (x, y, z) \in \left(2\mathbb{Z}\right)^{3}$ to denote positions of vertices and 
${\bf a} = (0,0,0), \, (1,1,0), \, (1,0,1), \, (0,1,1)$ to label the four FCC sublattices.
Hence, the set of $\{{\bf v}+{\bf a}\}$ represents the entire lattice sites with $\frac{1}{2}L^3$ spins for a linear size $L \in 2\mathbb{Z}$. Periodic boundary conditions (PBCs) are imposed.

The Hamiltonian is given by
\begin{align}\label{eq:TIM_model}
	H_{\rm TIM} = -\sum_{{\bf v}, {\bf a}^\prime} \left(J_+ \prod_{\bf a} s_{{(\bf v + {\bf a}^\prime )}+{\bf a}} + J_-\prod_{\bf a} s_{{(\bf v + {\bf a}^\prime )}-{\bf a}} \right),
\end{align}
which consists of four-body interactions living on elementary tetrahedra, and each site is shared by four $J_+$ and four $J_-$ tetrahedra.
We set the interaction to be isotropic as $J_+ = J_-=1$.

$H_{\rm TIM}$ is invariant under flipping all spins, including face centers, of an arbitrary $xy$, $yz$, or $zx$-layer.
This is easy to verify as either no or two spins in the interaction terms of Eq.~\eqref{eq:TIM_model} are affected.
There are in total $3L$ such plane-flip symmetries for a  lattice of size $L\times L \times L$. Specifically, we use $g_{i}^{x}$ to denote the symmetry of flipping spins in the $i^{\mathrm{th}}$ $xy$-layer, located at $x=i$. Similarly, $g_{j}^{y}$ and $g_{k}^{z}$ are used to denote the symmetries at $y=j$ and $z=k$, respectively. 
These symmetries are subject to three global constraints,
\begin{subequations} \label{eq:TIM_cons}
\begin{gather}
	\prod_{\text{odd }i}g_{i}^{x}=\prod_{\text{even }j}g_{j}^{y}\cdot\prod_{\text{even }k}g_{k}^{z}, \\
	\prod_{\text{odd }j}g_{j}^{y}=\prod_{\text{even }i}g_{i}^{x}\cdot\prod_{\text{even }k}g_{k}^{z}, \\
	\prod_{\text{odd }k}g_{k}^{z}=\prod_{\text{even }i}g_{i}^{x}\cdot\prod_{\text{even }j}g_{j}^{y}. 
\end{gather}
\end{subequations}
Thus, only $3L-3$ of them are independent, leading to a subextensive ground state degeneracy with  $\log_2{\rm GSD} = 3L-3$.

As two degenerate ground states differ by flipping at least $\frac{1}{2}L^2$ spins, i.e., an entire plane of the lattice, no finite-order perturbations can connect them at the thermodynamical limit.
Therefore, a long-range order spontaneously breaking the plane-flip symmetry remains permitted.

However, characterizations of such ordering are fundamentally distinguished from breaking a global symmetry.
As a necessary condition, the magnitude of a physical correlator or an order parameter needs to be preserved under arbitrary plane flips, which excludes any local order parameter.
Without losing generality, we consider the following minimal correlator
\begin{align} \label{eq:TIM_corr}
	G^z_{\rm TIM}(r) = \frac{8}{L^3} \sum_{\bf v} \mean{s_{\bf v} s_{{\bf v} + \hat{x} + \hat{y}} s_{{\bf v} + \hat{y} + r\hat{z}} s_{{\bf v} + \hat{x} + r\hat{z}}},
\end{align}
where the hat symbol indicates a unit vector and $r \in 2\mathbb{Z} +1$.

$G^z_{\rm TIM}(r)$ spans an irregular tetrahedron whose first and last pairs of spins belong to two different $xy$-layers separated by a distance $r$.
At the $r \rightarrow \infty$ limit, it relates to an order parameter $Q^z_{\rm TIM}$ as
$G^z_{\rm TIM}(r) \sim \left(Q^z_{\rm TIM}\right)^2$, with
\begin{align}\label{eq:TIM_op}
	Q^z_{\rm TIM} &= \frac{4}{L^3} \sum_{x, y} \mean{q^z_{\rm TIM}} \nonumber \\
	&= \frac{4}{L^3} \sum_{x, y}
	 \mean{\left|  \sum_{z} s_{\bf v} s_{{\bf v} + \hat{x} + \hat{y}} + s_{{\bf v} + \hat{y} + \hat{z}} s_{{\bf v} + \hat{x} + \hat{z}} \right|}.
\end{align}
Here, $q^z_{\rm TIM}$ can be viewed as an extended or semi-local ordering moment in an $xy$-layer, but itself is a {\it linewise} object that represents the total moments of local correlators $s_{\bf v} s_{{\bf v} + \hat{x} + \hat{y}}$ and $s_{{\bf v} + \hat{y} + \hat{z}} s_{{\bf v} + \hat{x} + \hat{z}}$ along an entire $z$-line of FCC unit cells. 
In view of dimensional reduction, it has a characteristic dimension $\dim(q^z_{\rm TIM}) = 1$ and a codimension ${\rm codim}(q^z_{\rm TIM}) = D-\dim(q^z_{\rm TIM}) = 2$. 
We hence refer to $Q^z_{\rm TIM}$ as a {\it sub-dimensional order parameter} to distinguish it from local order parameters and Wilson-loop order parameters.

One can similarly construct two equivalent order parameters $Q^x_{\rm TIM}$ and $Q^y_{\rm TIM}$. 
It is sufficient to use any of them.  

\subsection{Fractal Ising model} \label{sec:fractal_Ising}
The fractal Ising model is the dual of Haah's code~\cite{Vijay16}. Ising spins are placed at the vertices of a simple cubic lattice.
We again use ${\bf v} = \{x, y, z\}$ to represent vertex positions, but here ${\bf v} \in \mathbb{Z}^3$.
In addition, we define two finite sets of vectors ${\bf a}_1 \in \left\{(0,0,0), \, (1,0,0), \, (0,1,0), \, (0,0,1)\right\}$ and ${\bf a}_2 \in \left\{(0,0,0), \, (1,1,0), \, (1,0,1), \, (0,1,1)\right\}$ to label an arbitrary spin $s_{\bf v}$ and three of its nearest and next-nearest neighbors, respectively, without double counting the neighbors.

The Hamiltonian consists of two four-body interactions on the tetrahedra specified by $\{{\bf a}_1\}$ and $\{{\bf a}_2\}$, as depicted in Fig.~\ref{fig:model},
\begin{align} \label{eq:FIM_model}
	H_{\rm FIM} = -\sum_{{\bf v}} \left(J_1 \prod_{{\bf a}_1} s_{{\bf v}+{\bf a}_1} + J_2\prod_{{\bf a}_2} s_{{\bf v}+{\bf a}_2} \right),
\end{align}
where one $s_{\bf v}$ participates four $J_1$ tetrahedra and four $J_2$ tetrahedra.
For simplicity, we take $J_1 = J_2 = 1$.

In correspondence of Haah's code, $H_{\rm FIM}$ may have a fractal symmetry under PBCs.
There is no simple algebraic expression of its fractal symmetry generators and ground states~\cite{Note_generator}.
However, they can be constructed systematically by the approach provided in Sec.~\ref{sec:fractal_symmetry}.

The fractal symmetry leads to more exotic order parameters.
We can define invariant correlators by isotropically scaling the two interaction terms in  $H_{\rm FIM}$,
\begin{subequations}\label{eq:FIM_corr}
\begin{align}
	G_{\rm FIM} (r) & = \frac{1}{L^3} \sum_{\bf v}\mean{s_{\bf v} s_{{\bf v} + r\hat{x}} s_{{\bf v} + r\hat{y}} s_{{\bf v} + r\hat{z}} } \\
	G^{\prime}_{\rm FIM} (r)& = \frac{1}{L^3} \sum_{\bf v} \mean{s_{\bf v} s_{{\bf v} + r\hat{x} + r\hat{y}} s_{{\bf v} + r\hat{y} + r\hat{z}} s_{{\bf v} + r\hat{z} + r\hat{x}}}
\end{align}	
\end{subequations}
for $r=2^n$ with $n=1,2,\cdots$.
Both $G_{\rm FIM} (r)$ and $G'_{\rm FIM} (r)$ measure the long-range correlations of spins at the four corners of tetrahedra. 

We will discuss in Sec.~\ref{sec:fractal_symmetry} that locations of spins transformed by a symmetry operation depend on choices of fractal generators.
As a consequence, neither local ordering nor semi-local ordering like $Q^z_{\rm TIM}$ can be well defined.
This property also relates to the lack of string operators in Haah's code~\citep{Haah11, BravyiHaah11} whose locally creatable excitation patterns correspond to the interaction terms of $H_{\rm FIM}$~\cite{Haah13}.

Nonetheless, $G_{\rm FIM} \neq 0$ and $G'_{\rm FIM} \neq 0$ at $r \rightarrow \infty$ indicates a long-range order, and $G_{\rm FIM} = G'_{\rm FIM} = 1$ is only possible for ground states.
Hence, both $G_{\rm FIM}$ and $G'_{\rm FIM}$ can serve as order parameters to describe the fractal symmetry breaking of $H_{\rm FIM}$.
We refer to these non-local correlators as {\it fractal order parameters} to distinguish them from  sub-dimensional order parameters that admit a semi-local or dimension-reduction interpretation.

The non-local fractal order parameter is also essentially distinct from Wilson loop correlators in gauge theory.
Wilson loops generally vanish; deconfined and confined phases are distinguished by their decaying behaviors, namely the perimeter law and the area law, instead of the expectation values~\cite{Kogut79}.   
In contrast, in the fractal symmetry breaking phase, $G_{\rm FIM}$ and $G'_{\rm FIM}$ can robustly maintain a finite value at an arbitrarily large distance.

\begin{figure*}[!t]
	\includegraphics[width=0.9\textwidth]{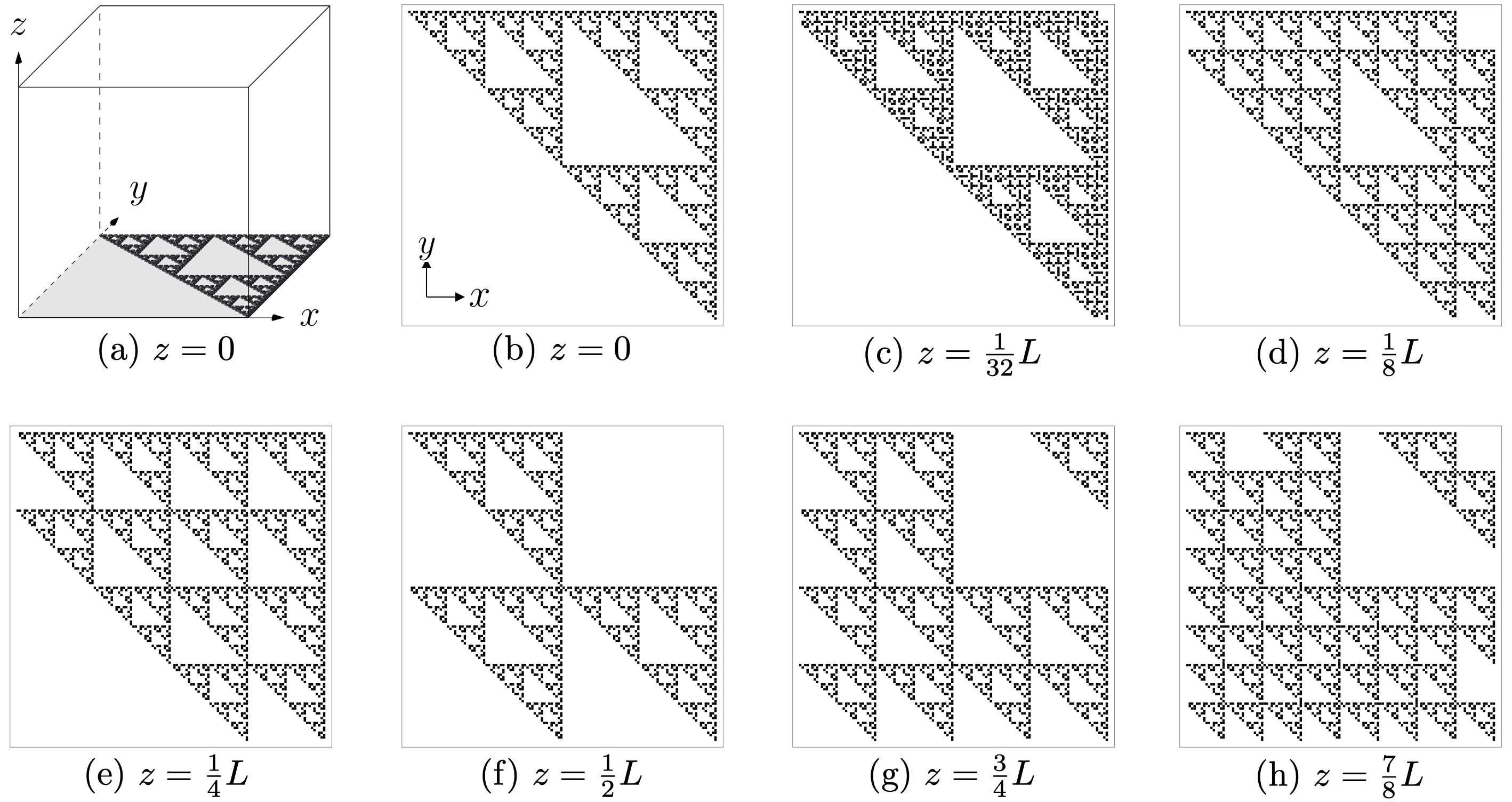}
	\caption{Fractal symmetry $X\left(f\right)$ that corresponds to $\left(c_{0},c_{1}\right)=\left(1,1\right)$ in a periodic system of linear size $L=128$. The pattern of flipped spins (black pixels), encoded by $f$, is illustrated by several layers in the $z$-direction. 
	(a, b) 3D and 2D representations for the pattern of spin flips in the $xy$-layer at $z=0$. (c-h) 2D representations for patterns of flipped spins in some other $xy$-layers; the presented $xy$-layers are located at $z=\frac{1}{32}L, \frac{1}{8}L, \frac{1}{4}L, \cdots, \frac{7}{8}L.$}
	\label{fig:fractal_sym}
\end{figure*}

\subsection{Self-duality} \label{sec:self-dual}
From the viewpoint of quantum error correction, self-duality of the tetrahedral and fractal Ising models reflects the equivalent role of Pauli X and Z errors in the checkerboard and Haah's codes.
At the level of spin models, this self-duality can also be understood directly through the Kramers-Wannier duality~\cite{Kramers41a, Kramers41b}.

The dual of $H_{\rm TIM}$ is given by placing a dual Ising spin $s_{{\bf v}^\star+ {\bf a}}$ at the center of an orignal $J_+$ tetrahedron $\prod_{\bf a} s_{{\bf v}+{\bf a}}$.
It is convenient to work with the shaded and empty unit cubes in Fig.~\ref{fig:model}, which are exclusively occupied by $J_+$ and $J_-$ tetrahedra, respectively.  
The dual lattice remains an FCC lattice with a uniform shift of $\frac{1}{2} (\hat{x}, \, \hat{y}, \hat{z})$; namely, the dual vertices and face centers are labeled by ${\bf v}^\star+ {\bf a} = {\bf v}+ {\bf a} + (\frac{1}{2}, \, \frac{1}{2}, \, \frac{1}{2})$. 
An original spin $s_{{\bf v}+ {\bf a}}$ has four neighboring shaded cubes, whose centers span a dual $J^\star_-$ tetrahedral $\prod_{\bf a} s_{{\bf v}^\star - {\bf a}}$.
Correspondingly, centers of the four empty neighboring cubes of $s_{{\bf v}+ {\bf a}}$ span a dual $J^\star_+$ tetrahedral $\prod_{\bf a} s_{{\bf v}^\star + {\bf a}}$.
Thus, the dual Hamiltonian preserves the form of $H_{\rm TIM}$, while the shaded and empty cubes are swapped on the dual side.

The dual of $H_{\rm FIM}$ can be analyzed analogously.
We again place a dual Ising spin $s_{{\bf v}^\star}$ at the center of an original cube. 
The dual lattice is a simple cubic lattice with vertices given by
$\left\{{\bf v}^\star \vert {\bf v}^\star = {\bf v} + (\frac{1}{2}, \, \frac{1}{2}, \, \frac{1}{2})\right\}$.
The two tetrahedron interactions in Eq.~\eqref{eq:FIM_model} are dual to 
$\prod_{{\bf a}_1} s_{{\bf v}^\star - {\bf a}_1}$ and $\prod_{{\bf a}_2} s_{{\bf v}^\star - {\bf a}_2}$, respectively, where the reverse sign of ${\bf a}_1$ and ${\bf a}_2$ merely swaps the incidence of spins in each unit cube.
Hence, the fractal Ising model is also self-dual. 

In virtue of this self-duality, we immediately have
\begin{align}
	& Z\left(\beta\right) \coloneqq \sum_{\{s\}} e^{-\beta H} = \sum_{\{s\}} \prod_i \omega(\beta, h_i) \nonumber \\
	& \qquad \propto \sum_{\{s^\star\}} \prod_i \omega^\star(\beta^\star, h^\star_i) = \sum_{\{s^\star\}} e^{-\beta^\star H^\star} \eqqcolon Z\left(\beta^\star\right). 
\end{align}
Here, $\omega^{(\star)} = e^{-\beta^{(\star)} h^{(\star)}_i}$ denotes the (dual) Boltzmann factor of a single interaction term $h^{(\star)}_i$, and $i$ is a shorthand label distinguishing those individual tetrahedra in (dual) $H_{\rm TIM}$ or $H_{\rm FIM}$.
As $h^{(\star)}_i = \pm 1$ is a binary function, we can rewrite the corresponding Boltzmann factors as $\omega_{\beta, n} = e^{\beta \cos{n\pi}}$ 
and $\omega^\star_{\beta^\star, k} = e^{\beta^\star \cos{k\pi}}$, with $n, k = 0, 1$.

The original and dual Boltzmann factors are related by a discrete Fourier transform,
$\omega_{\beta, n} = \frac{1}{\sqrt{2}} \sum_k  \omega^\star_{\beta^\star, k} e^{-i n k \pi}$, namely,
\begin{gather}
	\omega_{\beta, 0} = \frac{1}{\sqrt{2}} \big( \omega^\star_{\beta^\star, 0} + \omega^\star_{\beta^\star, 1}  \big) \\
	\omega_{\beta, 1} = \frac{1}{\sqrt{2}} \big( \omega^\star_{\beta^\star, 0} - \omega^\star_{\beta^\star, 1}  \big). \label{eq:weight_FT2}
\end{gather}  
Given the self-duality and provided they only have a single phase transition, which is the case for both of $H_{\rm TIM}$ and $H_{\rm FIM}$ (Sec.~\ref{sec:transition}), $Z\left(\beta\right)$ and $Z\left(\beta^\star\right)$ shall experience the same singularity.
Hence, at the transition point $\beta_c = \beta^\star_c$, we have 
\begin{align} \label{eq:weight_relation}
	\frac{\omega_{\beta_c, 0}}{\omega_{\beta_c, 1}} = \frac{\omega^\star_{\beta_c, 0}}{\omega^\star_{\beta_c, 1}},
\end{align}
whose solution gives the self-dual point $\beta_c = \frac{1}{2}\ln(\sqrt{2}+1)$.

\section{Fractal symmetry \& polynomial ring representation} \label{sec:fractal_symmetry}
We now discuss the symmetry of the fractal Ising model $H_{\mathrm{FIM}}$ utilizing a polynomial ring formulation.
This formalism is introduced in Ref.~\cite{Haah13} for describing translational invariant spin Hamiltonians and offers a convenient way to characterize fractal symmetries.
Before proceeding to compute the fractal symmetry and GSD of $H_{\mathrm{FIM}}$, we include a brief review of this formalism.

\subsection{Polynomial ring formalism}

The polynomial ring formalism is set up through a group ring $R=\mathbb{Z}_{2}\left[\Lambda\right]$, with $\Lambda=\{x^{i}y^{j}z^{k}|i,j,k\in\mathbb{Z}_{L}\}$ denoting the lattice translation group.
Namely, the coordinate of each vertex in the lattice is represented in a multiplicative notation, and PBCs are imposed by identifications $x^{L}=y^{L}=z^{L}=1$.
The ring $R$ can be viewed as a set consisting of all polynomials of the form 
\begin{equation}
f =\sum_{(i,j,k) \in\Lambda} a_{ijk}x^{i}y^{j}z^{k}.
\end{equation}
The coefficients $a_{ijk} = 0, 1 \in \mathbb{Z}_{2}$ are integers modulo $2$.
Intuitively, a polynomial $f \in R$ represents a collection of vertices or lattice vectors whose coefficients are $a_{ijk} = 1$.

For example, the sets $\left\{\mathbf{a}_{1}\right\}$ and $\left\{ \mathbf{a}_{2}\right\}$ in Eq.~\eqref{eq:FIM_model} holding the four-spin interactions correspond to the following two polynomials
\begin{equation}
\varepsilon_{1}=1+x+y+z \quad \text{and} \quad\varepsilon_{2}=1+xy+yz+zx.
\end{equation}

Furthermore, we use $X \left(f\right)$ to denote the operation of flipping spins on the set of vertices specified by $f$.
In the group ring formalism, the excitation pattern resulting from $X(f)$ acting on a ground state can be specified by an $R$-linear map from $R$ to $R^2$, which takes the form 
\begin{equation}\label{eq:error}
\varphi:f \mapsto\varphi\left(f\right)= \left(\overline{\varepsilon}_{1} f, \, \overline{\varepsilon}_{2} f\right).
\end{equation}
%The two components of $\varphi\left(f\right)= \left(\overline{\varepsilon}_{1} f, \, \overline{\varepsilon}_{2} f\right)$ describe the excitations associated with the two terms in Eq.~\eqref{eq:FIM_model} respectively. 
Here, standard polynomial multiplications are employed, and 
\begin{equation}\label{eq:epsilon}
\overline\varepsilon_{1}=1+ \overline{x} + \overline{y}+\overline{z}\quad\mathrm{and}\quad\overline\varepsilon_{2} = 1+ \overline{x}\overline{y}+\overline{y}\overline{z}+\overline{z}\overline{x}
\end{equation}
are the spatial inversion of $\varepsilon_{1}$ and $\varepsilon_{2}$, respectively.
We use the notations $\overline{x}\equiv x^{-1}$, $\overline{y}\equiv y^{-1}$,
and $\overline{z}\equiv z^{-1}$ for brevity.
The $R$-linearity in Eq.~(\ref{eq:error}) is a consequence of translation symmetry.

Physically, the polynomial pair $\left(\overline{\varepsilon}_{1} f, \, \overline{\varepsilon}_{2} f\right)$ describes the excitation pattern of the two interaction terms in $H_{\rm FIM}$.
This can be understood by considering a simplest polynomial $f=1$.
In this example, $\varphi(1) = \left(\overline{\varepsilon}_{1}, \, \overline{\varepsilon}_{2}\right)$ labels the locations of the excited  $J_1$ and $J_2$ tetrahedra in $H_{\rm FIM}$ due to a single spin flip at $x^0 y^0 z^0$.

\subsection{Polynomial representation of spin-flip symmetries}

Operation $X\left(f\right)$,  which flips the spins on the set of vertices specified by $f$, is a symmetry if and only if it creates no excitations, namely, $\varphi\left( f \right) =0$.
Therefore, the kernel of $\varphi$, 
\begin{equation}
	\ker\varphi\coloneqq\left\{ f\in R\;|\;\varphi\left(f\right)=0\right\}, 
\end{equation}
represents the group of spin flip symmetries of $H_{\mathrm{FIM}}$.

Explicitly, $\ker \varphi$ is specified by finding all the solutions $f \in R$ that satisfy
$\overline\varepsilon_1 f = \overline\varepsilon_2 f=0$.
As we show in App.~\ref{app:ring}, the first equation $\overline\varepsilon_1 f = 0$ implies that $f$ can be expressed as 
\begin{equation}\label{eq:rb}
f = b\,(1+tz+\left(tz\right)^{2}+\cdots+\left(tz\right)^{L-1}).
\end{equation}
Here, $t\equiv1+\overline{x}+\overline{y}$ for brevity. The factor $b\in\mathbb{Z}_{2}\left[\overline{x},\overline{y}\right]/(\overline{x}^{L}-1,\overline{y}^{L}-1)$ is a polynomial in $\overline{x}$ and $\overline{y}$, subject to the constraint
\begin{equation}\label{eq:bt}
b\,(t^{L}-1)=0.
\end{equation}

The form of $b$ is further solved from $\overline\varepsilon_2 f = 0$.
For convenience, we introduce 
\begin{align}\label{eq:alpha}
\overline{\alpha} & \coloneqq\left(\overline{x}+\overline{y}\right)\overline\varepsilon_1 +\overline\varepsilon_2=\overline{y}^{2}+\left(\overline{x}+1\right)\overline{y}+\overline{x}^{2}+\overline{x}+1.
\end{align}
Clearly, $\overline{\alpha} f = 0$ is equivalent to $\overline\varepsilon_2 f = 0$, provided $\overline\varepsilon_1 f = 0$ is satisfied.  
Nevertheless, $\overline{\alpha}$ is easier to handle as in Eq.~\eqref{eq:alpha} the dependence on $\overline{z}$ has been eliminated.

We next treat $b$ as a polynomial in $\overline{y}$ with coefficients in $\mathbb{Z}_{2}\left[\overline{x}\right]/(\overline{x}^{L}-1)$.
In order to find all the solutions of $\overline\alpha f = 0$, we temporarily lift the $y^L = 1$ PBC. 
Then, $\overline{\alpha} f =0$ reduces to
$\overline{\alpha}b=\left(c_{1}\overline{y}+c_{0}\right)(\overline{y}^{L}-1)$.
Thus, $b$ can formally be expressed as a polynomial long division
\begin{equation}\label{eq:b}
b=\frac{\left(c_{1}\overline{y}+c_{0}\right)(\overline{y}^{L}-1)}{\overline{\alpha}}
\end{equation}
and the division should produce zero reminder. Here, the coefficients $c_0$ and $c_1\in \mathbb{Z}_{2}\left[\overline{x}\right]/(\overline{x}^{L}-1)$ are polynomials in $\overline{x}$; two terms are  needed because the degree of $\overline{\alpha}$ in $\overline{y}$ is $2$.

The polynomial long division in Eq.~\eqref{eq:b}, where PBC $y^L=1$ is lifted, allows for the following physical interpretation. It ensures that $X(b)$ (namely, the operation that flips the set of spins specified by $b$ in the $xy$-plane) generates $\alpha$-type excitations only at the two open boundaries $y^0$ and $y^L$. Further, to ensure that excitations at both the $y^0$ end and the $y^L$ end get canceled when we reconnect the $y$-PBC, the polynomial long division has to produce zero remainder. The excitation patterns at both ends are described by $c_{1}\overline{y}+c_{0}$. It uniquely determines the form of $b$ by Eq.~\eqref{eq:b}.

There is no simple explicit formula of $b$~\cite{Note_generator}, and, in practice, it is more convenient to compute it on the fly.
Specifically, one can pick up two independent polynomials in $\overline{x}$ as the $c_0$ and $c_1$ coefficients and examine if $c_{1}\overline{y}+c_{0}$ is divisible by $\overline{\alpha}$ and if Eq.~\eqref{eq:bt} is satisfied.
In Sec.~\ref{sec:FIM_GSD} we discuss that the choices of $c_0$ and $c_1$ are straightforward in certain situations.

\subsection{Fractal symmetry \& ground state degeneracy} \label{sec:FIM_GSD}

A most distinctive feature of the fractal Ising model is that its symmetry group strongly depends on the system size $L$.
In particular, only specific $L$s support fractal symmetries under PBC.
As a consequence, the ground state degeneracy, which is determined by the number of spin-flip symmetries, i.e., ${\rm GSD} = \left|\ker\varphi\right|$, is also size-dependent.

The simplest choice of $L$s leading to a fractal symmetry is $L = 2^n$, with $n \in \mathbb{Z}^+$.
In such a case, Eq.~\eqref{eq:bt} is automatically fulfilled since coefficients of $f \in R$ are $\mathbb{Z}_2$ valued, hence $t^L = (1 + \overline{x} + \overline{y})^L \equiv 1$.  
The choices of $c_{0}$ and $c_{1}$ are then only subject to the requirement that 
$\left(c_{1}\overline{y}+c_{0}\right)(\overline{y}^{L}-1)$ is divisible by $\overline{\alpha}$. 
We demonstrate in App.~\ref{app:ring} that, for $L=2^n$, this remaining constraint also reduces to a much simpler version: $c_{0}+c_{1}$ containing an even number of terms, denoted $\left|c_{0}+c_{1}\right| = 0 \ {\rm mod} \ 2$.

As an example, we can choose $c_0 = c_1= 1$, which leads to a fractal pattern depicted in Fig.~\ref{fig:fractal_sym}.
Specifically, $c_0 = c_1 = x^0$ serves as a seed for Eq.~\eqref{eq:b}.
The resultant $b$ represents the pattern of flipped spins in the $xy$-layer at $z=0$.
And other layers are specified accordingly by those higher degree terms in Eq.~\eqref{eq:rb}.

By taking other combinations of $c_0$ and $c_1$, we can readily generate different fractal patterns and also global symmetries.
For instance, $c_0 = c_1 = \sum^{L-1}_{i=0} x^i$ leads to the obvious global symmetry which flips all spins in the entire lattice.

To determine the GSD, it is sufficient to know the number of spin-flip symmetries.
For $L=2^n$, it is simply the number of  choices for $c_0$ and $c_1$, under the constraint that $\left|c_{0}+c_{1}\right|$ is even. 
Thus, the GSD for this class of system sizes is $\left|\ker\varphi\right|=2^{2L-1}$.

When $L \neq 2^n$, the fractal Ising Hamilton realizes different symmetry groups and degeneracies.
It is not yet clear whether a simple expression exists to describe 
the ground state degeneracy for all values of $L$.
Nevertheless, we are able to enumerate several situations that cover typical systems sizes 
\begin{align}\label{eq:GSD}
\text{GSD}(H_{\mathrel{\text{FIM}}}) = \left|\ker\varphi\right| = 
\begin{cases}
2^{2L-1}, & L=2^{n},\\
2^{2L-5}, & L=4^{n}-1,\\
2^{2^m-1}, & L= 2^{m-1}(2^{n}+1), \\
2^{2^m-1}, & L= 2^{m-1}(2^{2n-1}-1), \\
\end{cases}
\end{align}
with integers $n, m \geq1$.
The first two classes in Eq.~\eqref{eq:GSD} are subextensive degeneracies due to different fractal symmetries; both have an exponent linear in $L$.
The latter two classes offer various lattice sizes (by changing $n$) that support a constant degeneracy for each fixed $m$. For instance, $L = 3, 5, 7, ...$ ($m=1$) have ${\rm GSD}=2$ and only the obvious global symmetry. Meanwhile, $L = 6, 10, 18, ...$ ($m=2$) give ${\rm GSD}=8$, indicating the presence of sophisticated global symmetries.  

We notice that the GSD in Eq.~\eqref{eq:GSD} equals the square root of the GSD for Haah's code~\cite{Haah13}.
This is because the fractal Ising model is associated with one type of qubit error (either Pauli $X$ or $Z$) in Haah's code. 
Hence, two copies of $H_{\rm FIM}$ make up the degeneracy of Haah's code.
In fact, a generalized version of this relation can be proved for general topological Calderbank-Shor-Steane (CSS) codes, which we will present in a separate work.

\begin{figure}[t]
	\centering
	\includegraphics[width=0.48\textwidth]{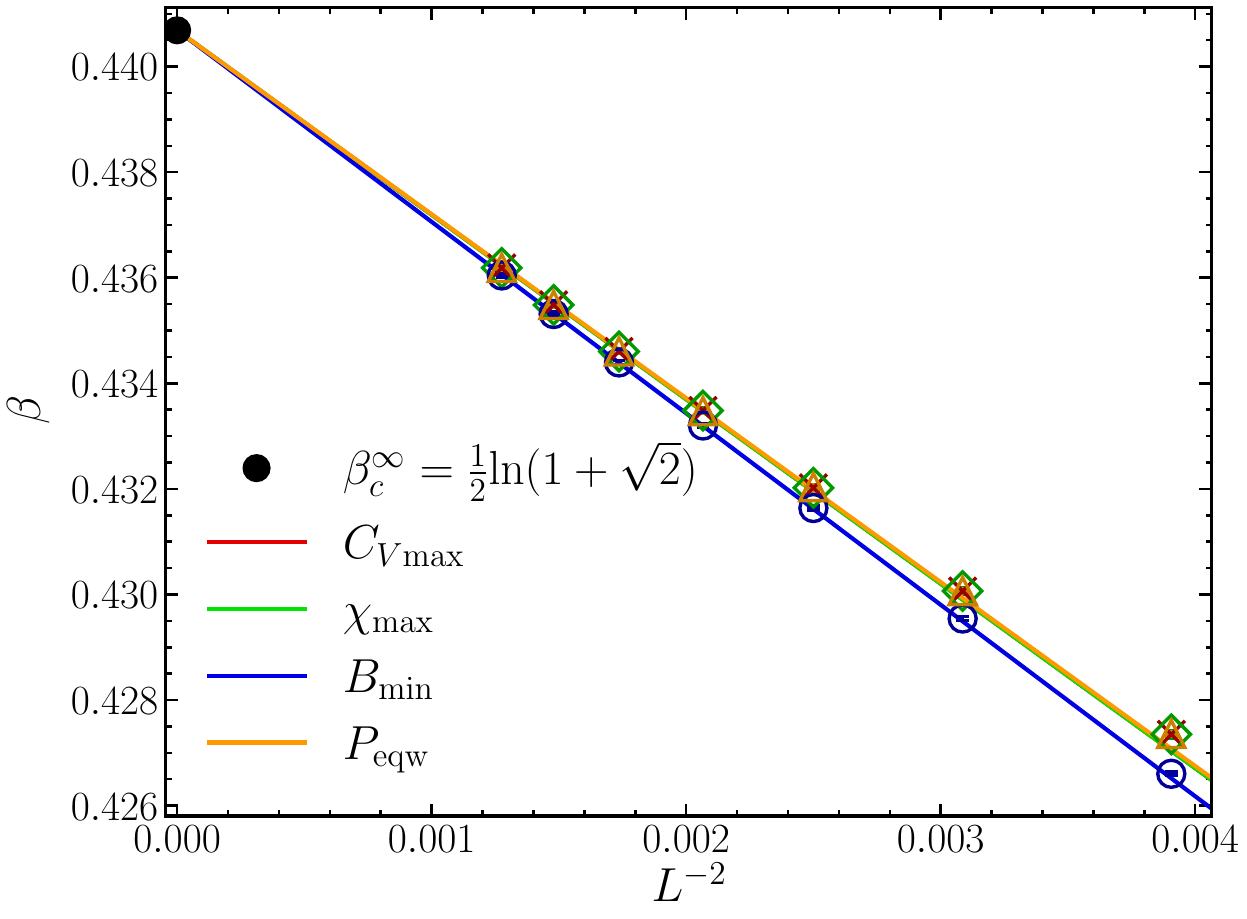}
	\caption{Anomalous scaling for the tetrahedral Ising model. Simulations are performed on lattices with $L = 16, 18, ..., 28$ ($N = \frac{1}{2}L^3$). Finite-size transition temperatures $\beta_c(L)$ are determined from the extrema of specific heat $C_v$, susceptibility $\chi$, and Binder cumulant $B$ and the equal-weight peaks of energy histogram $P(E)$. The thermodynamical $\beta_c^{\infty}$ is fixed by self-duality. All fits collapse onto the expected $\frac{1}{L^2}$ scaling.}
	\label{fig:TIM_scaling}
\end{figure}

\begin{figure*}[t]
	\centering
	\includegraphics[width=1\textwidth]{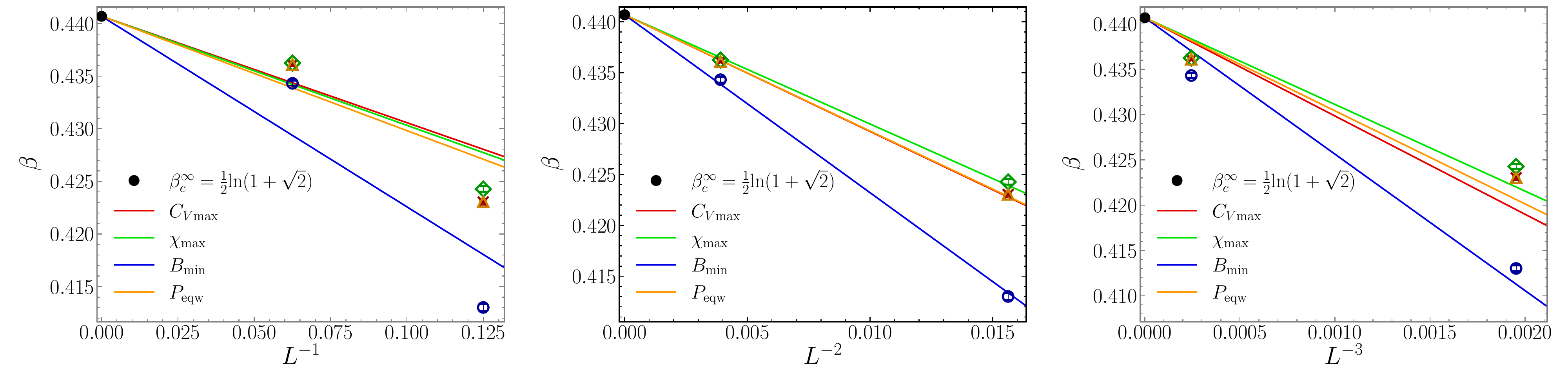}
	\caption{Anomalous scaling for the fractal Ising model, with $L = 8, \, 16, \, \infty$. The minimal three-point fit clearly prefers the expected $\frac{1}{L^2}$ scaling (middle), and excludes the a conventional $\frac{1}{L^3}$ fit (left) and a reference $\frac{1}{L}$ fit (right).}
	\label{fig:FIM_scaling}
\end{figure*}

\section{Phase transitions and scaling} \label{sec:transition}
Having established the symmetries, order parameters, and degeneracies of the two fracton spin models, we next discuss their phase transitions and scaling behaviors.
Numerical simulations are crucial as self-duality alone tells neither the number of phases nor the nature of their phase transitions. 

We perform large-scale Monte Carlo simulations by jointly utilizing heat-bath updates and a multi-canonical method~\cite{Janke08}, and PBCs are considered. 
Both $H_{\rm TIM}$ and $H_{\rm FIM}$ turn out to experience a very strong first-order phase transition.
The transition point in such situations is notoriously hard to locate precisely due to large hysteresis~\cite{BookBinder}. 
We overcome this problem with multi-canonical simulations by learning a nearly flat histogram.
Equilibrations are then ensured by the convergence of physical observables within statistical error bars.
See App.~\ref{app:simulation} for algorithm details and simulation parameters.

Phase transitions are determined by cross-checking the behaviors of the energy histogram $P(E)$, specific heat $C_V$, susceptibility $\chi$, Binder cumulant $B$, where
\begin{align}
& P(E) = \left\langle \delta\left(E-E'\right)\right\rangle , \label{eq:histogram} \\
& C_V = \frac{\beta^2}{N}\left( \left\langle E^{2}\right\rangle - \left\langle E \right\rangle^2 \right), \\
& \chi = \frac{\beta}{N}\left( \left\langle O^{2}\right\rangle - \left\langle O \right\rangle^2 \right), \\
& B =  1 - \frac{\mean{O^4}}{3\mean{O^2}^2}. \label{eq:binder}
\end{align} 
Here, $P(E)$ measures the distribution of total energy $E$ at a given temperature, $O \in \{Q^{z}_{\rm TIM}, \, G_{\rm FIM} \}$ denotes the sub-dimensional and fractal order parameters of the two models, and $N$ is the number of spins in the system.
The first-order nature of the phase transitions is concluded from diverging double peaks in histograms and negative dips in Binder cumulants (App.~\ref{app:simulation}).
Finite-size transition points $\beta_c(L)$ are located by the extrema of $C_V$,  $\chi$ and $B$ and equal-weighted peaks of $P(E)$.

In a standard first-order phase transition, $\beta_c(L)$ is known to satisfy a scaling relation
\begin{align}~\label{eq:scaling_standard}
	\beta_c(L) - \beta_c^{\infty} \sim \frac{\ln {\rm \Omega}}{E_o-E_d}+\mathcal{O}\big(\frac{1}{L^{D-1}}\big).
\end{align}
This relation can be derived from a two-phase model by assuming the system has an $\Omega$-degenerate ordered phase and a non-degenerate disordered phase, with an energy jump $E_d - E_o \sim L^D$ between the two phases~\cite{Lee91, Janke03}.
At a first-order phase transition, the ordered ($W_o$) and disordered ($W_d$) phases are expected to have the same weight, namely,
\begin{align}\label{eq:weights}
	\frac{W_o}{W_d} = \frac{\Omega e^{-\left( \beta E_o - \ln\Omega\right)}}{e^{-\left(\beta E_d - \ln 1\right)}} = 1.
\end{align}
Eq.~\eqref{eq:scaling_standard} is obtained by a series expansion of $\ln{\frac{W_o}{W_d}}$ around $\beta^\infty_c$.

However, in the current problem, the above scaling is subject to modification of subextensive degeneracies.
It is first demonstrated in studies of a plaquette Ising model (PIM) and an anisotropically coupled Ashkin-Teller model (ACATM).
Up to the leading order, the modification amounts to factoring out the size dependence in GSD~\cite{Mueller14, Johnston17}.
The consequent scaling thus turns into an anomalous one
\begin{align}\label{eq:scaling}
	\beta_c(L) \sim \beta_c^{\infty} + \frac{b}{L^{D-d}}+\mathcal{O}\big(\frac{1}{L^{D-d-1}}\big),
\end{align}
where $b$ is a non-universal constant prefactor, and $d$ reflects the power of $\log_2{\rm GSD} \sim L^{d}$.

Eq.~\eqref{eq:scaling} also indicates stronger finite-size effects in SDSB phase transitions, as the leading-order correction converges slower than in an usual first-order transition.
Nevertheless, for $H_{\rm TIM}$ and $H_{\rm FIM}$, once we know by numerics there is only a single phase transition, the thermochemical transition point  $\beta^\infty_c = \frac{1}{2}\ln\left(1+\sqrt{2}\right)$ is fixed by self-duality.
We are then left with a one-parameter fitting at the leading order.

We expect $H_{\rm TIM}$ to display an $\frac{1}{L^2}$ scaling as in the cases of PIM and ACATM~\cite{Mueller14}, as these 3D models all have a plane-flip symmetry.
This is indeed confirmed by our simulations by simulating large systems up to $L=28$.
As shown in Fig.~\ref{fig:TIM_scaling}, the transition temperatures computed using distinct estimates all firmly collapse onto a $\frac{1}{L^2}$-line.

It may be less obvious at first thought for $H_{\rm FIM}$.
However, considering the degeneracies given in Eq.~\eqref{eq:GSD}, this fractal symmetric model should also exhibit a $\frac{1}{L^2}$ scaling in the sequence of system sizes $L=2^n$ (or $L=4^n-1$).
Because of the constraints, only three lattice sizes $L \in \{8, \, 16, \, \infty\}$ are available in the $L=2^n$ sequence; others are either too small for a meaningful fit or too large for practical simulations.
In Fig.~\ref{fig:FIM_scaling}, we compare the $\frac{1}{L^2}$ fit with a standard $\frac{1}{L^3}$ fit and a reference $\frac{1}{L}$ fit.
It is remarkable that such a minimal three-point fitting neatly identifies the expected non-standard scaling.

In the other non-trivial sequence $L = 4^n-1$, the only accessible size is $L=15$ as the next relevant one is already $L=63$.
Nonetheless, given the agreement between theory and simulations so far, we presume that this class also falls into a $\frac{1}{L^2}$ scaling but converges to $\beta^\infty_c$ with a different slope due to the non-universal prefactor in Eq.~\eqref{eq:scaling}.

The results of $H_{\rm FIM}$ and  $H_{\rm TIM}$ indicate that, at the leading order, the anomalous first-order scaling depends only on the sub-extensive part in GSD, not specifically on their symmetry generators and global constraints.
In fact, SDSB phase transitions without a fracton relevance can also show the same scaling behavior as in the case of a hybrid symmetry breaking~\cite{Canossa23}.
However, order parameters in these situations are intrinsically different as discussed in Sec.~\ref{sec:models}.

We conclude this section by briefly commenting on the system size sequences $L=2^{m-1}(2^n+1)$ and $L=2^{m-1}(2^{2n-1}-1)$ in Eq.~\eqref{eq:GSD}. 
In these cases, if we keep $m$ fixed, the fractal Ising model has distinct global-symmetry generators and constant degeneracies at different system sizes.
A strong first-order phase transition is detected for all cases, but fitting the scaling is again  very resource-consuming, as one has to group them according to their degeneracy classes and resort to large system sizes.
However, given their constant degeneracies, one naturally expects them to fall into a conventional $\frac{1}{L^3}$ scaling but converge to the thermodynamic limit with different slopes.

\section{Summary} \label{sec:sum}
Fracton systems provide novel schemes of fault-tolerant quantum computation and unconventional states of matter and call for new phenomenologies.
In this work, we carried out a comprehensive study on two representative self-dual fracton spin models: the tetrahedral Ising model (TIM) and the fractal Ising model (FIM) that are the ungauging correspondence of checkerboard code and Haah's code, respectively.
We constructed their order parameters and analyzed their ground-state degeneracies and phase transitions. 

For the tetrahedral Ising model, its planar-flip symmetries lead to order parameters built from sub-dimensional linewise moments, as in Eq.~\eqref{eq:TIM_op}. 
The long-range order of the system emerges from correlations of these extended objects, instead of local correlators.
It represents a new type of order parameter, distinguished from those of Landau-types and Wilson loops. 
The construction may be generalized to general type-I fracton models that admit a foliation structure~\cite{Ma17, Shirley18} and also non-fracton sub-dimensional symmetric models that can be partitioned into coupled lines or layers~\cite{Canossa23}.

For the fractal Ising model, its fractal symmetry excludes the presence of local and semi-local ordering moments.
Instead, the long-range symmetric correlator serves as an effective order parameter, as in Eq.~\eqref{eq:FIM_corr}.
Namely, the system develops a long-range symmetry-broken order without ordering moments.
This sharply contrasts with the lack of local order parameters in symmetry-unbroken topological phases and may be viewed as a characteristic property of fractal symmetry breaking. 
Furthermore, as there are no simple expressions of fractal generators, the utilization of new algebraic tools, such as polynomial rings, becomes necessary for investigating such systems.

The phase transitions of both models belong to the regime of sub-dimensional symmetry breaking (SDSB) that features a sub-extensive number of degenerate ground states and long-range orders with non-local order parameters.
Such phase transitions appear to be commonly first-order in three dimensions.
Aside from the two models studied here, other examples include the dual spin models of the X-cube code~\cite{Mueller14, Johnston17} and two sub-dimensional symmetric spin models without a fracton correspondence~\cite{Canossa23}.
They are also in line with the first-order quantum phase transitions of perturbed X-cube~\cite{Devakul18}, checkerboard~\cite{Machaczek23}, and Haah's codes~\cite{Muhlhauser20}; namely, all three representative fracton codes when their excitations are condensed in the simplest manner.
Therefore, understanding the origin of these first-order transitions or finding exceptions will be an interesting exploration.

Despite being first-order, these phase transitions display an anomalous finite-size scaling, including the fractal Ising model (Sec.~\ref{sec:transition}).
The exponent of the scaling reflects the size dependence of the subextensive GSDs.
Thus, this anomalous scaling represents a common feature for fracton spin models and other sub-dimensional symmetric models. 

It would also be interesting to investigate the $Z_N$ forms of these fracton spin models.
Such $Z_N$ generalizations are a canonical topic in studying models with global symmetries and local symmetries.
In particular, $N$'s value may affect the structure of the underlying phase diagram and the nature of the associate phase transitions.
For instance, the $N$-state clock models~\cite{Oshikawa00, Hove03} and $Z_N$ lattice gauge theories~\cite{Bhanot80, Liu15} in three dimensions have a single continuous phase transition for any finite $N$, which extrapolate to the $3D$ $XY$ model and a confined $U(1)$ gauge theory when $N\rightarrow\infty$, respectively~\cite{Kogut79,Savit80}.  
However, the $N$-state clock models in two dimensions~\cite{Froehlich81, Ortiz12, Li20} and the $Z_N$ lattice gauge theories four dimensions~\cite{Horn79, Drouffe83} have an intermediate phase with algebraically decayed correlations for $N\gtrsim 5$, and consequently two separate phase transitions.
The appearance of such intermediate phases is deeply related to the self-duality of $2D$ spin models and $4D$ gauge theories~\cite{Ortiz12,Drouffe83}.
One can analogously ask whether self-dual $3D$ fracton spin models, particularly the $Z_N$ generalizations of the tetrahedral and fractal Ising models, could also support an intermediate phase at some larger $N$.

The breaking of a sub-dimensional symmetry constitutes a different type of phase transition than breaking a global symmetry or a gauge symmetry.
Our work advances the development of a complete understanding of SDSB phase transitions.
Moreover, the order parameters and scaling established here can further incorporate effects of quenched disorders and provide a crucial guide to studying the error resilience of the checkerboard code and Haah's code.

\begin{acknowledgments}
\emph{Acknowledgments}. G.C., K.L., and L.P. acknowledge support from FP7/ERC Consolidator Grant QSIMCORR, No.~771891, and the Deutsche Forschungsgemeinschaft (DFG, German Research Foundation) under Germany's Excellence Strategy -- EXC-2111 -- 390814868.
M.A.M.-D. acknowledges support from grants MINECO/FEDER Projects, PID2021-122547NB-I00 FIS2021, the ``MADQuantumCM'' project funded by Comunidad de Madrid. 
M.A.M.-D. has been financially supported by the Ministry of Economic Affairs and Digital Transformation of the Spanish Government through the QUANTUM ENIA project call---Quantum Spain project, and by the European Union through the Recovery, Transformation and Resilience Plan---NextGenerationEU within the framework of the Digital Spain 2026 Agenda.
M.A.M.-D. has also been partially supported by the U.S.Army Research Office through Grant No.~W911NF-14-1-0103.
H.S. acknowledges support from the National Natural Science Foundation of China (Grant No.~12047503).
K.L. acknowledges support from the New Cornerstone Science Foundation through the XPLORER PRIZE, Anhui Initiative in Quantum Information Technologies, and Shanghai Municipal Science and Technology Major Project (Grant No. 2019SHZDZX01).
The project/research is part of the Munich Quantum Valley, which is supported by the Bavarian state government with funds from the Hightech Agenda Bayern Plus.

The data used in this work are available in Ref.~\cite{data}.
\end{acknowledgments}

\begin{appendix}

\begin{figure*}[t]
\centering
\includegraphics[width =.8\textwidth]{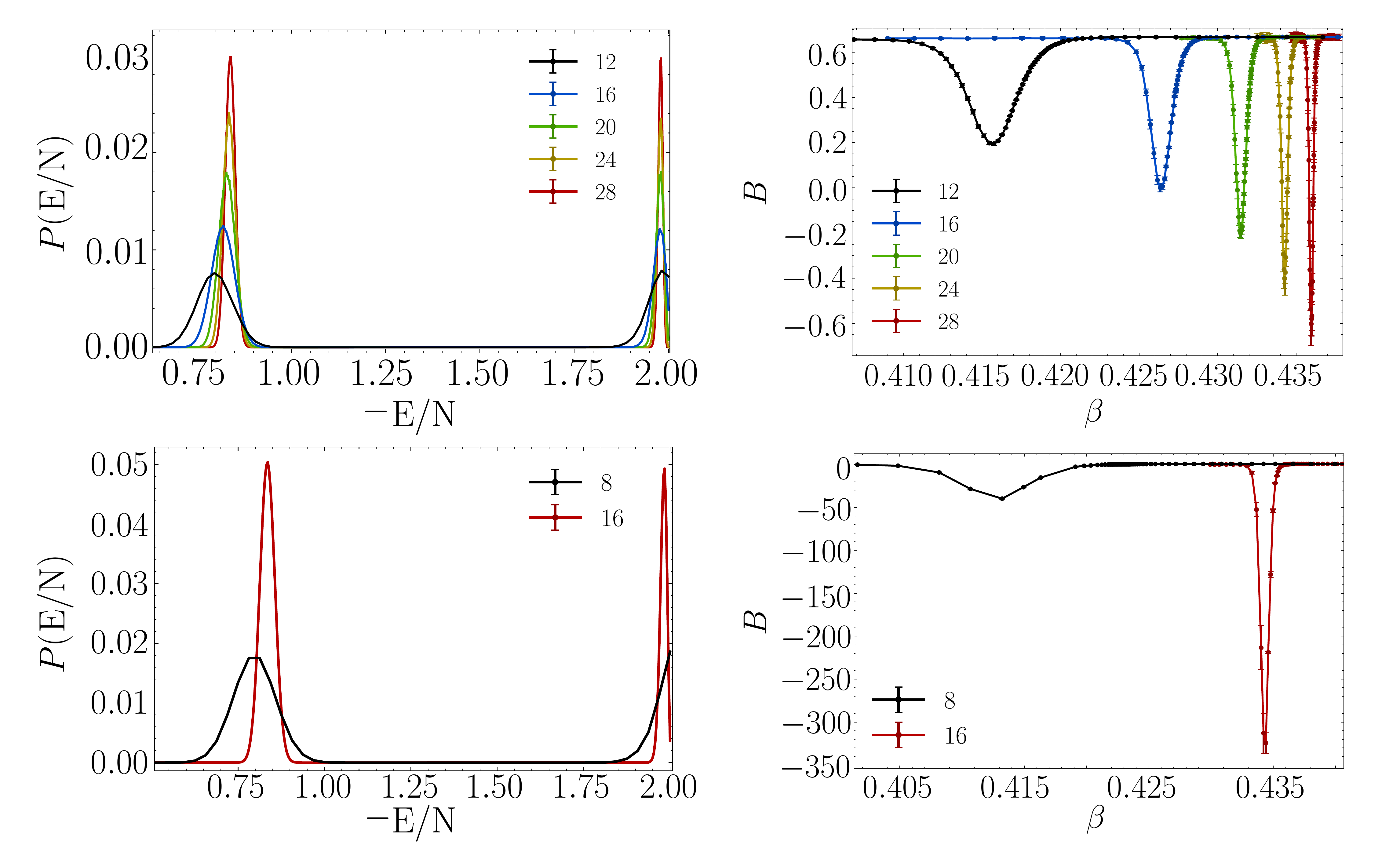}
\caption{Energy histograms (left column) and Binder cumulants (right column) for TIM (first row) and FIM (second row), computed with multicanonical simulations in the proximity of the phase transitions. The behavior of the dips in the Binder cumulants and the double peaks in the energy histograms for growing lattice sizes confirm the presence of a first-order phase transition in both models.}
 \label{fig:TIM-FIM_plots}
\end{figure*}

\section{Counting fractal symmetries}\label{app:ring}

We now show that $\overline\varepsilon_1 f = 0$ leads to Eq.~\eqref{eq:rb} and Eq.~\eqref{eq:bt}.
Formally, we can express $f$ as a polynomial in $z$, 
as $f = \sum_{k=0}^{L-1} b_k z^k$, whose coefficients
$b_{k}\in\mathbb{Z}_{2}\left[\overline{x},\overline{y}\right]/(\overline{x}^{L}-1,\overline{y}^{L}-1)$
are polynomials in $\overline{x}$ and $\overline{y}$. 
Then,
\begin{align}\label{eq:app_bz}
\overline\varepsilon_1 f & = (\overline{z}+t)\sum_{k=0}^{L-1}b_{k}z^{k} 
=\sum_{k=-1}^{L-2}b_{k+1}z^{k}+\sum_{k=0}^{L-1}b_{k}t\,z^{k} \nonumber\\
& =\sum_{k=0}^{L-2}\left(b_{k+1}+b_{k}t\right)z^{k}+\left(b_{0}+b_{L-1}t\right)z^{L-1},
\end{align}
where in the last line we used the PBC $\overline{z}=z^{L-1}$.

Clearly, $\overline\varepsilon_1 f = 0$ requires all coefficients in Eq.~\eqref{eq:app_bz} vanish.
Or equivalently,
\begin{align}
b_{0} & =b_{L-1}t,\label{eq:app_b0} \\
b_{k+1} & =b_{k}t,\forall k=0,1,\cdots,L-2. \label{eq:app_bk}
\end{align}
Eq.~\eqref{eq:app_bk} implies $b_{k}=t^{k}b$, where we set $b_0 \equiv b$ for brevity.
Hence, Eq.~\eqref{eq:rb} is obtained,
\begin{equation}
f = b(1+tz+\left(tz\right)^{2}+\cdots+\left(tz\right)^{L-1}). \nonumber
\end{equation}
Moreover, Eq.~\eqref{eq:app_b0} requires $b=bt^{L-1}t$, which is just Eq.~\eqref{eq:bt}.

We further prove that, for $L=2^n$, $\left(c_{1}\overline{y}+c_{0}\right)(\overline{y}^{L}-1)$ is divisible by $\overline\alpha$ if and only if $c_0+c_1$ contains an even number of terms,
where $c_{0}$ and $c_{1}\in\mathbb{Z}_{2}\left[\overline{x}\right]/(\overline{x}^{L}-1)$ are polynomials in $\overline{x}$. 

For convenience, we define new variables $p=\overline{x}+1$ and $q=\overline{y}+1$.
Accordingly, Eq.~\eqref{eq:alpha} becomes  
\begin{equation}\label{eq:app_alpha}
\overline{\alpha} = q^{2}+pq+p^{2}.
\end{equation}
%We are interested in system sizes $L=2^n$ below. In this situation, we have $p^{L} = \overline{x}^{L}-1$ and $q^{L} = \overline{y}^{L}-1$.

The condition that $\left(c_{1}\overline{y}+c_{0}\right)(\overline{y}^{L}-1)$ is divisible by $\overline\alpha$ can be expressed as
\begin{equation}\label{eq:app_pq}
\left(c_{1}\left(q+1\right)+c_{0}\right)q^{L}\equiv0 \pmod{\overline\alpha}.
\end{equation}
For system sizes $L=2^n$, we have $p^{L} = \overline{x}^{L}-1$ and $q^{L} = \overline{y}^{L}-1$. In this situation, $c_0$ and $c_1$ can be viewed as elements of $\mathbb{Z}_{2}\left[p\right]/(p^{L})$, with the $x$-PBC  represented as $p^L = 0$.

Notice the following identities 
\begin{align}
	q^{2^{n}} & \equiv(pq+p^{2})^{2^{n-1}}\equiv p^{2^{n-1}}q^{2^{n-1}}\equiv p^{2^{n-1}}(pq+p^{2})^{2^{n-2}}\nonumber \\
	& \equiv p^{2^{n-1}}(pq)^{2^{n-2}}\equiv p^{2^{n-1}+2^{n-2}}q^{2^{n-2}}\equiv p^{2^{n-1}+2^{n-2}+2^{n-3}}q^{2^{n-3}}\nonumber \\
	& \equiv\cdots\equiv p^{2^{n-1}+2^{n-2}+2^{n-3}+\cdots+1}q\equiv p^{2^{n}-1}q \pmod{\overline\alpha},\\
q^{2^{n}+1} & \equiv p^{2^{n}-1}q^{2}\equiv p^{2^{n}-1}(pq+p^{2})\equiv0 \pmod{\overline\alpha}.
\end{align}

Hence, for $L=2^n$, Eq.~\eqref{eq:app_pq} reduces to 
	\begin{equation}
	\left(c_{1}+c_{0}\right)p^{L-1}q\equiv0 \pmod{\overline\alpha},
\end{equation}
which holds if and only if $c_{1}+c_{0}$ is divisible by $p$, hence by $\overline{x}+1$ using the original variable $\overline{x}$. 

For simplification, the condition can be equivalently stated as requiring $c_{0}+c_{1}$ to contain an even number of terms. This equivalence can be demonstrated by observing the dichotomy $c_{1}+c_{0}\equiv0\;\text{or}\;1 \pmod{\overline{x}+1}$ and its correspondence with whether $c_{1}+c_{0}$ contains an even or odd number of terms.

\section{Multi-canonical simulations} \label{app:simulation}
As shown in Fig.~\ref{fig:TIM-FIM_plots}, both fracton spin models $H_{\rm TIM}$ and $H_{\rm FIM}$ showcase a strong first-order phase transition with a significant energy barrier $\Delta E = E_d - E_o$ between the disordered and ordered phase.
Such an energy barrier strongly suppresses intermediate states and makes tunneling events between the two phases unlikely.
As a result, canonical simulations will be trapped in one of the metastable states.
To overcome this issue, we performed multicanonical (MC) Monte Carlo simulations that modify the canonical probability distribution and promote the exploration of intermediate states.
We now summarize the main steps of the multicanonical algorithm and refer to Ref.~\cite{Janke08} for a detailed introduction.

The multicanonical algorithm aims to learn a flat energy distribution in a sufficiently large temperature interval covering $E_d$ and $E_o$.
To do so, we define the following multicanonical partition function
\begin{equation}
\label{eq:Z_MC}
Z_{\rm MC}(\beta) = \sum_{E} \rho(E) e^{-\beta E - g(\beta, E)},
\end{equation}
where $\rho(E)$ is the density of states at energy $E$, and $e^{- g(\beta, E)}$ is an unknown weighting factor to be learnt.
It is not hard to see that $Z_{\rm MC}(\beta)$ gives a flat energy distribution in the interested interval $[E_o, E_d]$ only if the weighting factors satisfying $e^{g(\beta, E)}=\rho(E) e^{-\beta E}$ for all energies therein.
This is achieved by initializing $g(\beta, E)=0 \,\forall\, E$ and updating them iteratively by repeating the following steps.
\begin{enumerate}
\item Run a set of Monte Carlo simulations in parallel at a given temperature $\beta$ near the phase transition with $N_R$ different replicas (independent initializations). We use the standard heat-bath algorithm for Monte Carlo updates, but the Boltzmann weights are now evaluated according to Eq.~\eqref{eq:Z_MC}.
	Compute the energy histogram $h_j(E)$ for the $j$-th replica; namely, keep tracking the frequency of each energy $E\in[E_o,E_d]$.
\item Once a previous set of Monte Carlo runs finishes, update $g(\beta, E)$ with an inversion rule
\begin{equation}
g(\beta, E) \rightarrow g(\beta, E) +  \ln \,H(E) - \left< \ln \, H(E')\right>_{E^\prime}.
\end{equation} 
Here, $H(E) = \sum_{j=1}^{N_R}h_j(E) / N_R$ is the averaging histogram at a given energy $E$, and $\langle \dots \rangle_{E^\prime}$ denotes a thermal average over the energy interval of interest.
\end{enumerate}

After each iteration, the new $g(\beta, E)$ will suppress those frequently visited states but promote the less visited states in the previous Monte Carlo runs.
Such processes are repeated until $H(E)$ satisfies a flatness condition $H(E)/\langle H(E') \rangle_{E^\prime} \sim 1$, $\forall\,E\in [E_o,E_d]$.
In practice, we consider $H(E)$ is flat if $\left|H(E)- \langle H(E') \rangle_{E^\prime} \right|/\langle H(E') \rangle_{E^\prime} \lesssim 0.1$ for three consecutive iterations.

\begin{figure}[t]
\centering
\includegraphics[width =.45\textwidth]{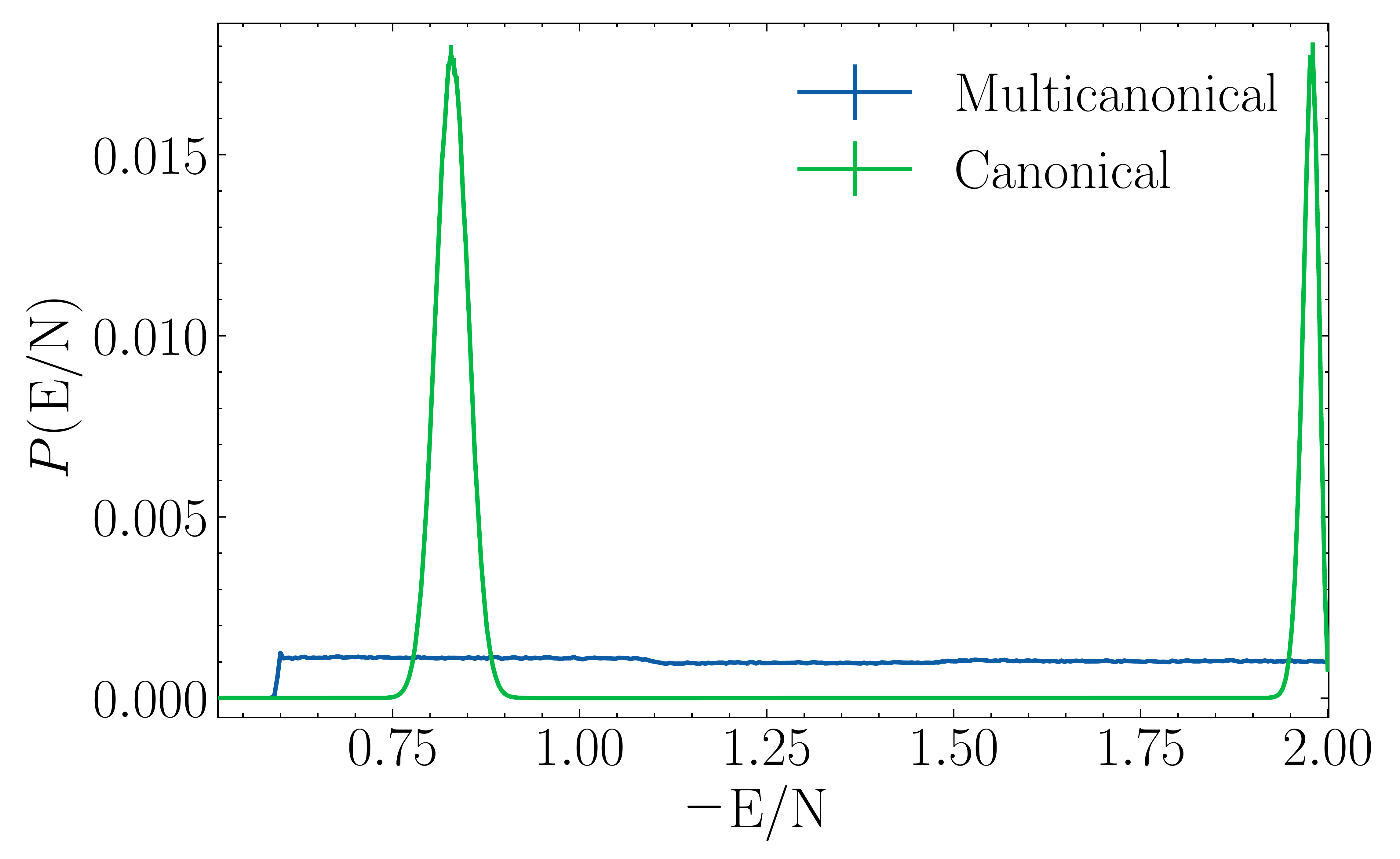}
  \caption{Energy histograms at the estimated transition point obtained from the multicanonical simulations and the resulting canonical reweighted distribution for the Tetrahedral Ising model at L=20.}
 \label{fig:MC-CA_E_hist}
\end{figure}

At the end of the above weight-learning procedure, we obtain $g(\beta, E)$ that gives us a nearly flat energy distribution $H_{\rm MC}(E)$ over $[E_o, E_d]$, as shown in Fig.~\ref{fig:MC-CA_E_hist}.
We can use the learned weight to adjust the Boltzmann factor at any nearby temperature $\beta^\prime$ by setting $g(\beta^\prime, E) = g(\beta, E) + (\beta - \beta^\prime) E$ and perform new Monte Carlo simulations according to Eq.~\eqref{eq:Z_MC} for measurement runs.
In general, $g(\beta^\prime, E)$ does not have a perfect flat energy distribution for $\beta^\prime \neq \beta$, but it is enough to efficiently explore the intermediate states in the interested interval $[E_o, E_d]$.

Physical observables are defined by canonical expectation values. Given the learned reweighting factors $e^{-g(\beta, E)}$, we can systematically derive the expectation value of an arbitrary observable $\langle \mathcal{O} \rangle$ using the following relation
\begin{flalign}
\langle \mathcal{O}\,e^{g(\beta, E)} \rangle_{\rm MC}&= \frac{1}{Z_{\rm MC}}\sum_{E^\prime} \mathcal{O}(E^\prime) \rho(E^\prime) e^{-\beta E^\prime - g(\beta, E')}e^{ g(\beta, E^\prime)} \nonumber\\
&= \left<\mathcal O\right> \frac{Z}{Z_{\rm MC}}.
\end{flalign}
The global normalization factor is $\frac{Z}{Z_{\rm MC}} = \sum_{E^\prime} H_{\rm MC}(E^\prime)\, e^{g(\beta, E^\prime)}$. 

In Table~\ref{tab:simulation_sweeps}, we summarize the simulation parameters.
It is worth noting that, as multicanonical simulations explore a larger configuration space than a canonical one, larger amounts of Monte Carlo updates are required to obtain accurate estimates of physical expectation values.

\begin{table}[tb]
\renewcommand{\arraystretch}{1.5}
\centering
\begin{tabular}{p{1.3cm} p{1.3cm} p{1.3cm} p{1.3cm} p{1.3cm}}
\toprule
\hline 
$L$ & $N_R$ & $N_{WL}$ & $N_{T}$ & $N_{S}$ \\
\midrule
\multicolumn{5}{l}{\bf Tetrahedral Ising model}\\
12 & 64 & $10^5$ & 64 & $5.5\times 10^7$  \\
14 & 64 & $10^5$ & 64 & $8.8\times 10^7$  \\
16 & 64 & $10^5$ & 64 & $10^8$  \\
18 & 64 & $5\times 10^5$ & 64 & $2.5\times 10^8$  \\
20 & 64 & $10^6$ & 64 & $2.9 \times 10^8$  \\
22 & 64 & $10^6$ & 64 & $5.5 \times 10^8$  \\
24 & 64 & $2\times 10^6$ & 64 & $7.7\times 10^8$  \\
26 & 64 & $2\times 10^6$ & 64 & $10^9$  \\
28 & 64 & $3\times 10^6$ & 64 & $10^9$  \\
\midrule
\multicolumn{5}{l}{\bf Fractal Ising model}\\
4 & 64 & $10^5$  & 64 & $4\times 10^6$  \\
8 & 64 & $10^5$  & 64 & $10^7$  \\
16 & 64 & $3\times 10^5$ & 64 & $1.2 \times 10^9$  \\
\hline
\bottomrule
\end{tabular}
\caption{Simulation parameters for the weight-learning process and the subsequent multicanonical simulations. $N_R$ denotes the number of independent simulations from which histogram averages are taken in the weight learning process. $N_{WL}$ is the number of sweeps used for each replica in the last three iterations.
 Multicanonical simulations are then run at $N_T$ different temperatures for $N_S$ number of Monte Carlo sweeps at each temperature.}
\label{tab:simulation_sweeps}
\end{table}

\end{appendix}

\bibliography{fracton.bib}

\begin{thebibliography}{84}
\bibitem [{not()}]{note_subdim}%
  \BibitemOpen
  \href@noop {} {}\bibinfo {note} {Such symmetries are also dubbed subsystem
  symmetry in the literature. As the term ``subsystem'' has different meanings
  in different contexts, for clarity, we instead use ``sub-dimensional'' to
  refer to entire sub-manifolds or fractals of a system.}\BibitemShut {Stop}%
\bibitem [{Not()}]{Note_generator}%
  \BibitemOpen
  \href@noop {} {}\bibinfo {note} {The symmetry generator for the fractal Ising
  model given in the appendix of Ref.~\cite{Vijay16} is a global
  symmetry.}\BibitemShut {Stop}%
\end{thebibliography}

\end{document}